\documentclass[twocolumn,aps,prd,superscriptaddress,floatfix]{revtex4-1}

\usepackage{amsmath,amsthm,amssymb}
\usepackage{graphicx}
\usepackage{color}
\usepackage{xcolor}
\usepackage{float}
\usepackage{multirow,array}

\usepackage[colorlinks,
bookmarks=true,
linkcolor=blue,
urlcolor=blue,
anchorcolor=black,
citecolor=blue
]{hyperref}

\colorlet{darkblue}{blue!60!black}

\begin{document}
\title{GW190814: Circumstantial Evidence for Up-Down Quark Star}
\author{Zheng Cao}
\affiliation{School of Physics and Astronomy, Shanghai Key Laboratory for Particle Physics and Cosmology, and Key Laboratory for Particle Astrophysics and Cosmology (MOE), Shanghai Jiao Tong University, Shanghai 200240, China}

\author{Lie-Wen Chen}
\thanks{Corresponding author}
\email{lwchen@sjtu.edu.cn}
\affiliation{School of Physics and Astronomy, Shanghai Key Laboratory for Particle Physics and Cosmology, and Key Laboratory for Particle Astrophysics and Cosmology (MOE), Shanghai Jiao Tong University, Shanghai 200240, China}

\author{Peng-Cheng Chu}
\affiliation{Science school, Qingdao University of technology, Qingdao 266000, China }

\author{Ying Zhou}
\affiliation{School of Physics and Astronomy, Shanghai Key Laboratory for Particle Physics and Cosmology, and Key Laboratory for Particle Astrophysics and Cosmology (MOE), Shanghai Jiao Tong University, Shanghai 200240, China}
\date{\today}

\begin{abstract}
Within a confining quark matter model which considers phenomenologically the quark confinement and asymptotic freedom as well as the chiral symmetry restoration and quark deconfinement at high baryon density,
we find that if the up-down quark matter ({\it ud}QM) is more stable than nuclear matter and strange quark matter (SQM), the maximum mass of static quark stars with {\it ud}QM is $2.87M_{\odot}$ under agreement with both the constraints on star tidal deformability from gravitational wave signal GW170817 and the mass-radius of PSR J0030+0451 and PSR J0740+6620 measured by NICER.
In contrast, the conventional strange quark star with the SQM that is more stable than nuclear matter while the nuclear matter is more stable than {\it ud}QM, has a maximum static mass of only $1.87M_{\odot}$ and its radius significantly deviates from NICER's constraint.
Our results thus provide circumstantial evidence suggesting the recently reported GW190814's secondary component with a mass of $2.59^{+0.08}_{-0.09}M_\odot$ could be an up-down quark star.
\end{abstract}

\maketitle

\section{Introduction}
The LIGO Scientific and Virgo Collaborations recently announced the gravitational wave (GW) event GW190814~\citep{Abb20}, a binary coalescence involving a primary black hole~(BH) with mass $23.2^{+1.1}_{-1.0}M_\odot$ and a secondary compact object with mass $2.59^{+0.08}_{-0.09}M_\odot$ ($90\%$~C.L.). The secondary falls into the so-called ``mass-gap'' between known neutron stars (NSs) and BHs~\citep{Bai98,Oze10,Far11,Oze12}.
Indeed,
a maximum mass of $M_{\rm{TOV}} \approx 2.3M_\odot$~\citep{ZhouY19a} for static NSs has been obtained based on
a careful investigation by applying a single density functional model to simultaneously analyze the data of finite nuclei and the NS tidal deformability limit from GW170817~\citep{Abb17a,Abb17b,Abb18,Abb19} together with the constraints on the equation of state (EOS) of symmetric nuclear matter (SNM) at suprasaturation densities from flow data in heavy-ion collisions (HICs)~\citep{Dan02}. We note that the relatively small value of $M_{\rm{TOV}} \approx 2.3M_\odot$ is mainly due to the soft EOS of SNM constrained by the flow data in HICs~\citep{Dan02}, and this is further confirmed recently by relativistic mean-field calculations~\citep{Fat20,Hua20}.

The $M_{\rm{TOV}} \approx 2.3M_\odot$ has also been obtained from analyzing the electromagnetic (EM) counterparts during the post-merger evolution of GW170817~\citep{Mar17,Rez18,Rui18,Shi19}, and the remnant of GW170817 with mass $\sim 2.7M_\odot$ has been suggested to form a BH
at the end.
On the other hand, it should be noted that,
due to the fact that the detectors are not sensitive enough, the EM observations of GW170817 do not provide definitive evidence for or against a long-lived NS as a possible post-merger outcome~\citep{Abb17c}.
In particular,
Ai et al.~\citep{Ai18} consider five
stiff EOSs with a range of $M_{\rm{TOV}}$ from $2.28$ to $2.75$ solar mass that could fulfill the tight limit on
tidal deformability of GW170817 and show that a millisecond NS with relatively low
dipole magnetic field could also meet the constraints from GW and multi-band EM observations
of GW170817.
Furthermore, as pointed out by Piro et.al~\citep{Pir19}, a long-lived NS is not only allowed, but also helpful to
interpret some of the data, such as blue kilonova, red kilonova component as well as the
spectral features of lanthanides elements.
Therefore, the EM observations of GW170817 alone cannot rule out the possibility of a long-lived massive NS as the remnant of GW170817.

Some studies indicate that
the $M_{\rm{TOV}}$ for static NSs can be
larger than about $2.7M_\odot$ while still satisfying the tidal deformability constraint of GW170817~\citep{Ann18}
or the combined observational constraints of
GW170817 and GW190814 (assuming it is a NS-BH merger)~\citep{Abb20},
however, they
ignore the flow data constraint on the EOS of
SNM at suprasaturation densities, which can give a very strong
limit on $M_{\rm{TOV}}$ for static NSs.
Therefore, the secondary is unlikely to be a non- or slowly-rotating NS under the constraint on
the EOS of SNM at suprasaturation densities from the flow data,
although it could be a rapidly-rotating NS~\citep{Abb20,Mos20,ZhangNB20,Tso20,Dex20,Sed20} but further understanding is needed on how an NS-BH system could merge before dissipating such extreme natal NS spin angular momentum~\citep{Abb20}. Some studies~\citep{Fat20,Ess20,Tew20} suggest that the secondary should be a BH.

Besides heavy rapidly-rotating NS or light BH, the quark star (QS) provides another candidate for the GW190814's secondary component.
According to the hypothesis by Bodmer~\citep{Bod71}, Terazawa~\citep{Ter79} and Witten~\citep{Wit84}, many studies~\citep{Bom04,Sta07,Her11} suggest that NSs may be converted to strange quark stars (SQSs) which are made purely of absolutely stable deconfined $u$, $d$, and $s$ quark matter~(QM) with some leptons, i.e., strange quark matter~(SQM).
There are arguments that the QS is probably unable to explain the pulsar glitches~\citep{Mad88} and the quasi-periodic oscillations~(QPOs) for the
highly magnetized compact stars~\citep{Wat07}. Because of the complicated structure of QM, however, alternative explanations for the pulsar glitches still exist~\citep{Jai06,Ang14} and also the observation of QPOs cannot conclusively rule out the QS hypothesis due to the unknown features of the stars~\citep{Mil19a,Ren20}.
A robust calculation for dense QM from {\it ab initio} QCD is still a big challenge and our knowledge on QM properties related to compact stars essentially relies on effective models~\citep{Web05,Bub14}.
In fact, some models predict that SQM may not be absolutely stable~\citep{Bub99,Wan03,Rat03,Kla15,Hol18,Wan19a}. Especially, a recent study~\citep{Hol18} with a phenomenological quark-meson model suggests that the up-down QM~($ud$QM) can be more stable than the ordinary nuclear matter and SQM, and accordingly the up-down quark star~($ud$QS) is explored~\citep{Wan19b,Zha19,Zha20,Ren20}.

Furthermore, we note that there are recently some studies~\citep{Lai18,Buc19} that investigate
the kilonova formation in a binary QS merger.
In particular, Lai {\it et al.}~\citep{Lai18} demonstrate that the light curve of the
kilonova AT 2017gfo following GW170817 could be explained by considering the decaying
strangeon nuggets and remnant star spin-down.
Bucciantini {\it et al.}~\citep{Buc19}
show that the evaporation of most quark matter fragments into
nucleons (mostly neutrons) takes place close to the central region of the merger and thus
the evolution of that material is similar to that of the nucleonic material ejected during the
merger, therefore it can contribute to the kilonova signal (and nucleosynthesis) in the same way.
These works
provide viability that a binary QS merger could produce sufficient neutron
rich ejecta for r-process nucleosynthesis and thus for kilonova.
Therefore, the GW170817 could be a binary QS merger with remnant as a massive QS.

In this work, we examine the possibility that the GW190814's secondary is a QS within a confining QM~(CQM) model~\citep{Dey98,Bag04} which considers phenomenologically some basic features of QCD and can reasonably describe baryon properties~\citep{Dey86,Bag04}.
We find while the GW190814's secondary cannot be a conventional SQS, where the SQM is more stable than nuclear matter while the nuclear matter is more stable than $ud$QM, it could be a $ud$QS, i.e. $ud$QM is more stable than the ordinary nuclear matter and SQM so the quark star is $ud$QS, under agreement with both the constraints on star tidal deformability from GW170817~\citep{Abb19} and the mass-radius~(M-R) of PSR J0030+0451~\citep{Ril19,Mil19b} and PSR J0740+6620~\citep{Mil21,Ril21} from NICER.

The paper is organized as follows. In Section~\ref{Sec2}, we give a brief description of the CQM model and its application in QS calculations. In Section~\ref{Sec3}, we present the results and discussions of our present work. The conclusion is given in Section~\ref{Sec4}.

\section{Model and method}
\label{Sec2}
In the CQM model~\citep{Dey98}, the Hamiltonian of quark matter is expressed as
\begin{equation}
\label{Eq:Hami}
H=\sum_{i}\left(\alpha_{i} \cdot \mathbf{p}_{i}+\beta_{i} M_{i}\right)+\sum_{i<j} \frac{\lambda(i) \cdot \lambda(j)}{4} V_{i j},
\end{equation}
where $i$ ($j$) represents quark flavor, $\alpha_i$ and $\beta_i$\ come from
Dirac equation, $\lambda$ is the color SU(3) matrix for interacting quarks,
$V_{ij}$ is the vector interaction between two quarks, and $M_i$ is the
quark mass related to quark scalar potential and chiral condensates~\citep{Dey98,Chu17}.
To mimic the chiral symmetry restoration at high baryon density,
the $M_i$ is parameterized as
\begin{equation}
M_{i}=m_{i}+m_i^*\operatorname{sech}(\nu_i n_{B}/n_{0}),
\label{Eq:qmass}
\end{equation}
where $m_{i}$ is the current quark mass, $m_i^*$ is a parameter determining the constituent quark mass in vacuum, $n_{B}$ is the baryon number density, $n_{0}=0.17~\mathrm{fm^{-3}}$ is normal nuclear matter density, and $\nu_k$ is a parameter controlling the density dependence of quark mass. One sees that the quark mass $M_{i}$ decreases with $n_{B}$ and the chiral symmetry is restored at high baryon density if the current quark mass $m_{i}$ can be neglected.

For the vector interaction $V_{ij}$, we adopt here the modified screened Richardson potential~\citep{Bag04}, i.e.,
\begin{eqnarray}
\label{Eq:Vpot}
 V_{ij}(\mathbf{q}^{2})&=&\frac{12 \pi}{27}\big[V_{\rm{AF}}(\mathbf{q}^{2})  + V_{\rm{CF}}(\mathbf{q}^{2})\big],
\end{eqnarray}
with
\begin{eqnarray}
 V_{\rm{AF}}(\mathbf{q}^{2})&=&\frac{1}{ (\mathbf{q}^{2}+m_g^{2})\ln \left(1+\frac{\mathbf{q}^{2}+m_g^{2}}{\Lambda^{2}}\right)}
 -\frac{\Lambda^{2}}{(\mathbf{q}^{2}+m_g^{2})^2}, \notag \\
 V_{\rm{CF}}(\mathbf{q}^{2})&=&\frac{\Lambda'^{2}}{(\mathbf{q}^{2}+m_g^{2})^2}, \notag
\end{eqnarray}
where $\mathbf{q} = \mathbf{k}_{i}-\mathbf{k}_{j}$ is the momentum transfer between the two interacting quarks, the gluon mass $m_g$ is introduced to describe the screening effects on the vector potential in medium due to pair creation and infrared divergence, $\Lambda$ represents the asymptotic
freedom~(AF) scale as the $V_{\rm{AF}}(\mathbf{q}^{2})$ goes asymptotically zero for large $\mathbf{q}^{2}$, and $\Lambda'$ corresponds to the confinement~(CF) scale as the $V_{\rm{CF}}(\mathbf{q}^{2})$ reduces to a linear confinement for small $\mathbf{q}^{2}$ (Note: $m_g^{2}$ vanishes in vacuum and $\lambda(i) \cdot \lambda(j) = -8/3$ for quark-quark interactions).
To the lowest order, the gluon mass $m_g$ is related to the screening length $D$ according to~\citep{Kap79}
\begin{equation}
\label{leng}
m_g^2 = \left(D^{-1}\right)^{2}=\frac{2 \alpha_{0}}{\pi} \sum_{i} k_{i}^{f} \sqrt{\left(k_{i}^{f}\right)^{2}+M_{i}^{2}},
\end{equation}
where $\alpha_{0}$ is the perturbative quark-gluon coupling constant and $k_{i}^{f}=\left(\pi^{2} n_{i}\right)^{1 / 3}$ is the quark Fermi momentum with $n_{i}$ being the quark number density. At high baryon density, the $m_g^{2}$ becomes large and the $V(\mathbf{q}^{2})$ approaches to zero, leading to quark deconfinement.

It is clear that the present CQM model phenomenologically incorporates four basic features of QCD, namely, asymptotic freedom and linear quark confinement as well as chiral symmetry restoration and quark deconfinement at high baryon density.
In the original Richardson potential for heavy quarks~\citep{Ric79}, one has $\Lambda'=\Lambda$ and thus the AF and CF have the same scale. However, while $\Lambda$ is $\sim 100$~MeV from perturbative QCD~\citep{Shi79,Dey98}, $\Lambda'$ is found to be $\sim 400$~MeV from heavy and light meson spectroscopies as well as baryon properties~\citep{Ric79,Cra84,Dey86} where the CF plays an important role. In particular, the values of $\Lambda=100$~MeV and $\Lambda'=350$~MeV are shown to successfully describe the energies and magnetic moments of $\Delta^{++}$ and $\Omega^{-}$~\citep{Bag04}.
In this work,
we adopt $\Lambda=100$ MeV, $\Lambda'=350$ MeV, $m_{u}=4$ MeV, $m_{d}=7$ MeV and $m_{s}=150$ MeV to be consistent with Ref.~\citep{Bag04}, and $\alpha_{0}=0.65$ following Ref.~\citep{Bag06}.
We also set $m_u^*=331$ MeV, $m_d^*=328$ MeV and $m_s^*=377$ MeV to match the vacuum constituent quark mass $M_{u 0}=M_{d 0}=335$~MeV and $M_{s 0}=527$~MeV from SU(3) Nambu-Jona-Lasinio~(NJL) model with the parameter set HK~\citep{Hat94}.
Furthermore, we assume $\nu_{u}=\nu_{d} \equiv \nu_{u d}$ for simplicity, and thus we have only
two free parameters, namely, $\nu_{ud}$ and $\nu_s$.
We note that using the isospin-dependent quark mass formula ($\nu_{u}\neq \nu_{d}$) with one more parameter can help to enhance the $M_{\rm TOV}$ by about $0.2M_\odot$ for SQSs~\citep{Chu17}.

For the details of QS calculations within the CQM model, the reader is referred to Refs.~\citep{Dey98,Chu17}. The dimensionless tidal deformability $\Lambda_k$ for a QS with mass $M_k$ and radius $R$ can be expressed as
$
\Lambda_k=\frac{2}{3}(R/M_k)^{5} k_{2}
$ where $k_2$ is the dimensionless quadrupole tidal Love number.
In a binary system, the mass weighted tidal deformability $\tilde{\Lambda}$ is defined as~\citep{Fla08,Hin08}
\begin{eqnarray}
\tilde{\Lambda}=\frac{16}{13} \frac{(12q+1)\Lambda_{1} + (12+q)q^4\Lambda_{2}}{(1+q)^5},
\end{eqnarray}
where $\Lambda_{1}$ ($\Lambda_{2}$) is for the component with mass $M_1$ ($M_2$) in the binary and $q=M_2/M_1 \le 1$ is the mass ratio. Special boundary condition~\citep{Hin10,Pos10} should be applied in determining $k_{2}$ since the self-bound QSs have a sharp discontinuity of energy density at the surface.
For all the QS calculations in this work, the causality condition is guaranteed, i.e., the sound speed $c_s \equiv \sqrt{d P/d \epsilon} \le 1$ where $P$ is the pressure and $\epsilon$ is the energy density of the QS matter. We use the natural units with $\hbar = c = G = 1$ in the present work.

\section{Result and discussion}
\label{Sec3}
We mainly consider two QS scenarios, namely, the conventional SQS and the $ud$QS. For the conventional SQS, SQM is absolutely stable by satisfying the so-called stability window~\citep{Far84}, i.e., the minimum energy per baryon $E_{\rm{min}}$ of SQM should be less than that of the observed stable nuclei (i.e., $930$~MeV) while the $E_{\rm{min}}$ of $ud$QM should be larger than $930$~MeV to be consistent with empirical nuclear physics. For $ud$QS, $ud$QM is absolutely stable by requiring the $E_{\rm{min}}$ of $ud$QM should be less than $930$~MeV and the $E_{\rm{min}}$ of SQM.
By varying the values of $\nu_{ud}$ and $\nu_{s}$,
we find the $M_{\rm{TOV}}$ for the conventional SQS reaches its maximum value of  $1.87M_\odot$ with $\nu_{ud} = 0.39$ and $\nu_{s} = 1.07$ (the parameter set is denoted as SQS1.87) while the $M_{\rm{TOV}}$ for $ud$QS reaches its maximum value of $3.30M_\odot$ with $\nu_{ud} = 1.04$ and $\nu_{s} \le 0.94$ (the parameter set is denoted as udQS3.30).
Therefore, the $ud$QS can have a much larger $M_{\rm{TOV}}$ than the conventional SQS, and the $M_{\rm{TOV}}$ of the conventional SQS is significantly smaller than the mass $2.59^{+0.08}_{-0.09}M_\odot$ from GW190814.

The self-bound QS with a larger $M_{\rm{TOV}}$ generally has a larger radius and thus a larger $\tilde{\Lambda}$ for a fixed mass. The GW170817 is the first confirmed merger event of two pulsars, and with minimum assumptions it puts a strong constraint of $\tilde{\Lambda} = 300^{+420}_{-230}$ ($90\%$ highest posterior density interval) with the binary mass ratio $q = 0.73$-$1.00$ and chirp mass $M_c = 1.186^{+0.001}_{-0.001}M_\odot$ for the low-spin prior~\citep{Abb19}.
Assuming GW170817 is a binary QS merger, we find the udQS3.30 strongly violates the constraint $\tilde{\Lambda} = 300^{+420}_{-230}$ as it predicts $\tilde{\Lambda}(q=0.73) = 1260$.
Since $\tilde{\Lambda}(q)$ generally increases with $q$ for a fixed $M_c$, the GW170817 constraint implies $\tilde{\Lambda}(q=0.73) \le 720$ and $\tilde{\Lambda}(q=1) \ge 70$ for $M_c = 1.186$.
To be consistent with the GW170817 constraint on $\tilde{\Lambda}$, we find the $ud$QS can have a maximum of $M_{\rm{TOV}} = 2.87M_\odot$ with $\nu_{ud} = 0.843$ and $\nu_{s} \le 0.72$ (the parameter set is denoted as udQS2.87), and the udQS2.87 predicts $\tilde{\Lambda}(q=0.73) = 720$ for $M_c = 1.186M_\odot$ and $\Lambda(1.4M_\odot) = 677$.

\begin{figure}[!htbp]
\centering
\includegraphics[width=0.95\linewidth] {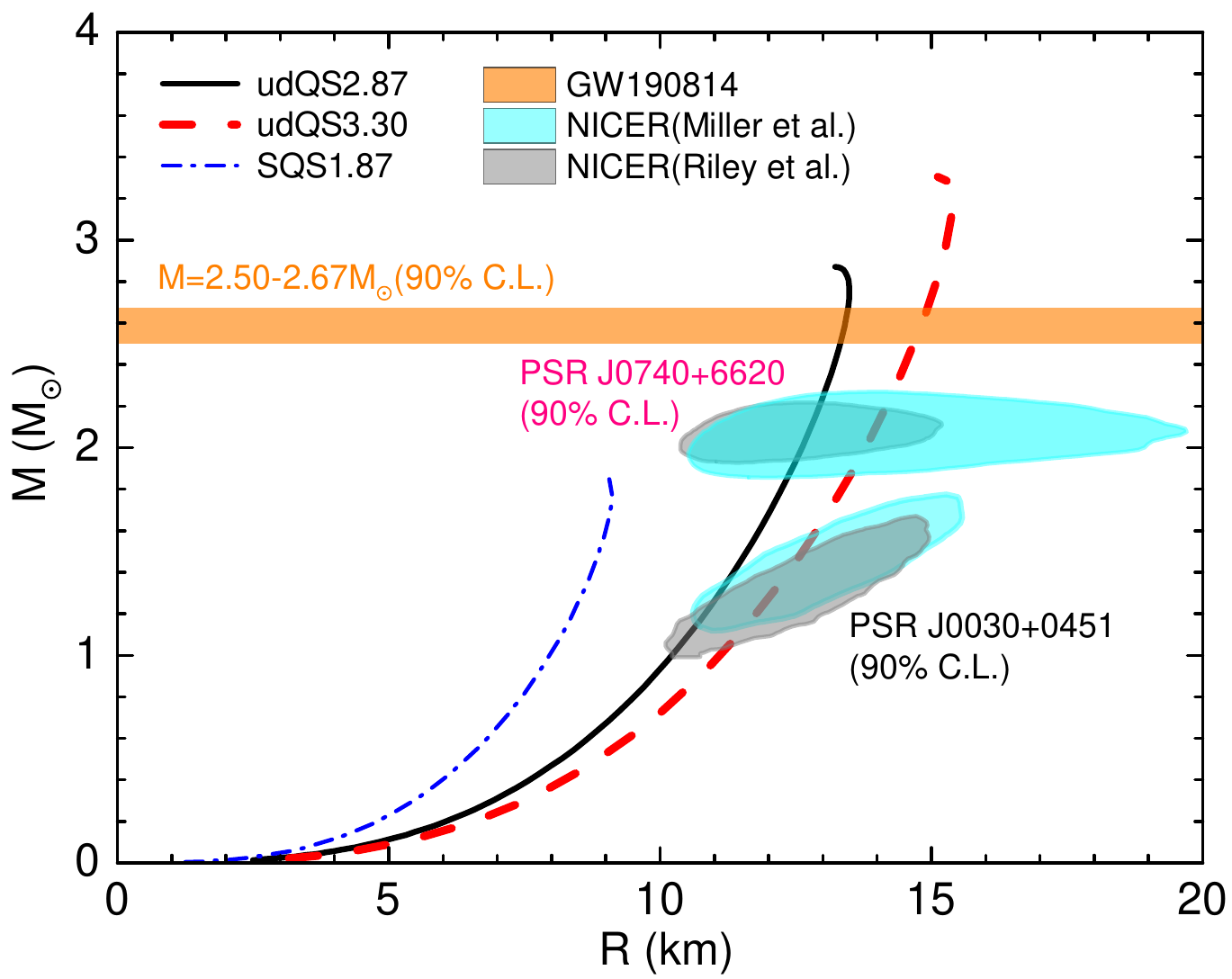}
\caption{M-R relation in the CQM model with SQS1.87, udQS2.87 and udQS3.30.
For comparison, the mass $2.59^{+0.08}_{-0.09}M_\odot$ ($90\%$~C.L.) from GW190814 (orange band),
and the two independent constraints for PSR J0030+0451~\citep{Mil19b,Ril19} and PSR J0740+6620~\citep{Mil21,Ril21} from NICER (90\% C.L.), are also included.}
\label{MR}
\end{figure}

Figure~\ref{MR} displays the QS M-R relation with SQS1.87, udQS2.87 and udQS3.30.
For comparison, we also include in Fig.~\ref{MR} the mass $2.59^{+0.08}_{-0.09}M_\odot$ from GW190814 and the two independent simultaneous M-R measurement from NICER by analyzing the X-ray data for the millisecond pulsar PSR J0030+0451~\citep{Ril19,Mil19b} with mass around $1.4M_\odot$ and PSR J0740+6620~\citep{Ril21,Mil21} with mass around $2.0M_\odot$.
It is seen that the result with udQS2.87 is in good agreement with NICER constraints on the M-R for PSR J0030+0451 and PSR J0740+6620 at $90\%$~C.L..
One also sees that the M-R of the conventional SQSs with SQS1.87 is significantly far from the NICER constraints. Therefore, our results suggest while the GW190814's secondary cannot be a conventional SQS, it could be a $ud$QS.

\begin{figure}[h!]
\centering
\includegraphics[width=0.95\linewidth]{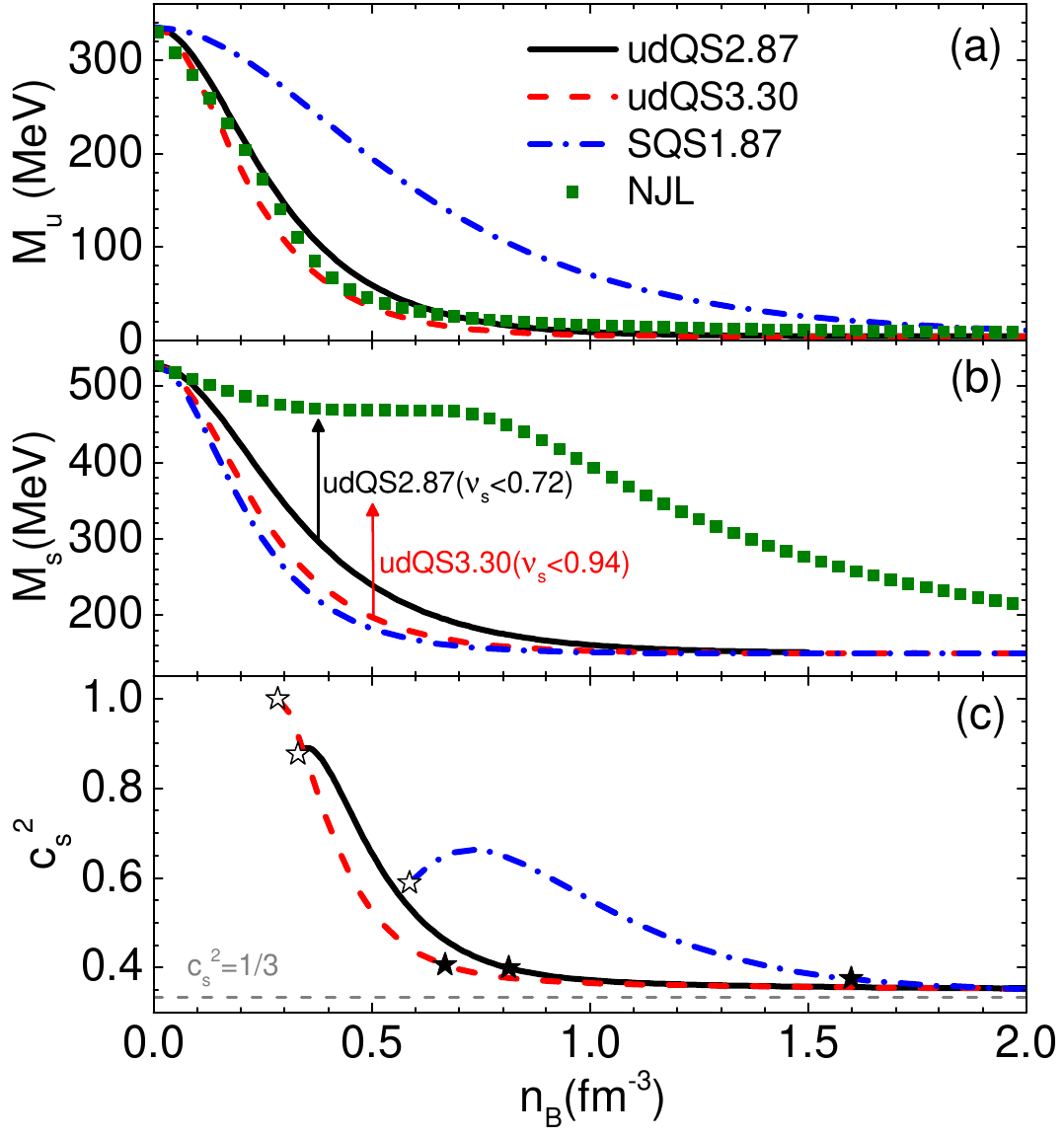}
\caption{The $u$-quark mass~(a), $s$-quark mass~(b) and the squared sound speed $c_s^2$ in QS matter~(c) as functions of the baryon density in the CQM model with SQS1.87, udQS2.87 and udQS3.30.
The corresponding $u$- and $s$-quark masses from NJL model are also plotted for comparison.
The solid (open) stars in panel~(c) indicate the center (surface) density of the maximum mass configuration of the QS.}
\label{Mu}
\end{figure}

It is instructive to see some implications of our results.
Shown in Fig.~\ref{Mu} is the density dependence of the quark mass $M_{u}$ and $M_s$ as well as the squared sound speed $c_s^2$ with SQS1.87, udQS2.87 and udQS3.30. The results of $M_{u}$ and $M_s$ from SU(3) NJL model with HK~\citep{Hat94} are also plotted for comparison.
The $M_d$ is not shown here since it is very similar to the $M_u$ except the slight difference in the current mass.
We note that
the NJL model is very successful to describe the properties of hadrons with QCD chiral symmetry with the model parameters fitted to the properties of mesons and the quark condensate in vacuum, and it provides an effective approach to describe the chiral symmetry restoration at high baryon densities. Therefore, the in-medium quark mass based on the well-known parameter set in the NJL model can be used as a guidance for the density dependent quark mass in the present CQM model, at least at lower densities where the linear approximation should be valid for the density dependence of the quark mass (and chiral condensates) (See, e.g., Refs.~\citep{Cai18,Cai19}). On the other hand, the NJL model cannot describe the quark confinement which is included in the present CQM model via the vector interactions $V_{ij}$ in Eq.~(\ref{Eq:Vpot}).
In the present CQM model,
the two free parameters $\nu_{ud}$ and $\nu_s$ determine the density dependence of the quark mass,
and larger values of $\nu_{ud}$ or $ \nu_s $ make the quark mass decay more rapidly with density and thus facilitate the chiral symmetry restoration at high density if the current quark mass is neglected.
It is remarkable to see from Fig.~\ref{Mu}(a) that while the $M_{u}$ with SQS1.87 significantly deviates from the NJL result, the udQS2.87 and udQS3.30 predicts a quite similar $M_{u}$ as the NJL model. Furthermore, the udQS2.87(udQS3.30) requires $\nu_{s} \le 0.72(0.94)$, and thus the corresponding $M_s$ is also in harmony with the NJL result as seen in Fig.~\ref{Mu}(b).
Our results thus indicate that the $u$, $d$ and $s$ quark masses with udQS2.87 and udQS3.30 nicely agree with the results from the NJL model.

The sound speed $c_s$ is an important quantity characterizing the properties of the matter, and especially it measures the stiffness of the matter EOS. From Fig.~\ref{Mu}(c), one sees that the $c_s$ in QS matter at lower densities is close to the light velocity in vacuum and significantly violates the so-called conformal bound $c_s^2 \le 1/3$~\citep{Bed15}, and the $c_s$ approaches to the conformal limit $\sqrt{1/3}$ at high density due to quark deconfinement, asymptotic freedom and the small quark masses.
We note that for the maximum mass configuration, the center baryon density $n_{\rm{B,c}}$ (surface baryon density $n_{\rm{B,surf}}$) of the QS is $1.60$~fm$^{-3}$ ($0.59$~fm$^{-3}$) for SQS1.87, $0.81$~fm$^{-3}$ ($0.33$~fm$^{-3}$) for udQS2.87, and $0.67$~fm$^{-3}$ ($0.28$~fm$^{-3}$) for udQS3.30. Due to the self-bound feature of the QS matter, the QS has a finite surface density $n_{\rm{B,surf}}$ where the pressure is zero and the $c_s$ is close to the light velocity in vacuum, and this special density dependence of $c_s$ with a peak value of $c_s \approx 1$ around the QS surface but $c_s \approx \sqrt{1/3}$ at the QS center leads to the QS can have a larger maximum mass. Furthermore, it is seen clearly from Fig.~\ref{Mu}(c) that a stiffer QM EOS generally leads to a maximum mass configuration with smaller $n_{\rm{B,c}}$ and $n_{\rm{B,surf}}$, and thus larger QS radii as seen in Fig.~\ref{MR}.

We further examine a special case in which SQM is more stable than $ud$QM while the latter is more stable than nuclear matter, i.e., the $E_{\rm{min}}$ of $ud$QM is less than $930$~MeV but larger than the $E_{\rm{min}}$ of SQM. The SQM that obeys this stability condition is dubbed as {\it unconventional} SQM.
We note such {\it unconventional} SQS made up by {\it unconventional} SQM can have a very large mass, e.g., the $M_{\rm{TOV}} = 5.6M_\odot$ with $\nu_{ud} = 3.0$ and $\nu_{s} = 2.7$ (the parameter set is dubbed as ucSQS5.6). However, the ucSQS5.6 predicts $\tilde{\Lambda}(q=0.73) = 11868$,
drastically violating the $\tilde{\Lambda}$ constraint from GW170817. To satisfy the $\tilde{\Lambda}$ extracted from GW170817, we find the $M_{\rm{TOV}}$ of the {\it unconventional} SQS can have a maximum of $2.80M_\odot$ with $\nu_{ud} = 0.96$ and $\nu_{s} = 1.08$ (the parameter set is denoted as ucSQS2.80), and in this case one has $\tilde{\Lambda}(q=0.73) = 720$ and $\Lambda(1.4M_\odot) = 675$.
However, we note the ucSQS2.80 predicts an extremely strong density dependence of $M_s$ with $\nu_{s} = 1.08$, which significantly deviates from the density dependence of the $M_s$, even at very low densities, based on the NJL model as shown in Fig.~\ref{Mu}(b), and thus is strongly disfavored.
We also note a number of calculations~\citep{Hor04,Kov09,Wei11,Rod11,Vas17} for SQM based on MIT-bag-like models or color-flavor-locked phase with typical values of the gap parameter indicate that the SQS $M_{\rm{TOV}}$ can be larger than $2.5M_\odot$, but the $\tilde{\Lambda}$ constraint from GW170817 is generally violated. These results suggest again that the GW190814's secondary cannot be a SQS.

There are a number of important implications if the $ud$QM is absolutely stable~\citep{Hol18,Zha20,Xia20,Iid20,Ren20}.
One interesting example is the existence of stable $ud$QM nuggets or $ud$lets, and consequently a new ``continent of stability'' of $ud$lets is expected to appear as an extension of the hypothesized ``island of stability'' around mass number $A\sim 300$ in nuclear landscape~\citep{Hol18}.
While a sophisticated calculation is needed to predict the properties of the $ud$lets, here we assume a $ud$let is a sphere of radius $R$ containing $u$~($d$) quarks with constant number density $n_u$~($n_d$), and the energy per baryon of $ud$lets with baryon number $A$ and electric charge number $Z$ can then be estimated by using a simple mass formula,
i.e., $E_{ud\rm{lets}}(A,Z) = E(n_u,n_d) + 4\pi \Sigma R^2/A + 3Z^2e^2/(5AR)$, where $E(n_u,n_d)$ is the energy per baryon of the $ud$QM and $\Sigma$ is the surface tension coefficient of the $ud$QM.
Currently, the $\Sigma$ is poorly known and its value largely depends on the model and method~\citep{Jai06,Wen10,Xia18,Xia20}. The $^{294}_{118}$Og~\citep{Oga12} is the heaviest nucleus discovered so far, and its energy per baryon is $931.975$~MeV~\citep{Wan17}. So the $E_{ud\rm{lets}}(A,Z)$ for $A=294$ and $Z=118$ should be larger than $931.975$~MeV to avoid the strong decay of $^{294}_{118}$Og into a $ud$let, and this leads to a constraint of $\Sigma \ge 152$~MeV/fm$^2$.
We also adopt Thomas-Fermi approximation~\citep{Xia20}, which considers scalar and vector potentials, to calculate the minimum surface tension that prevents $^{294}_{118}\mathrm{Og}$ from decaying into a $ud$let, and find $\Sigma \ge 165$~MeV/fm$^2$.
These estimates of $\Sigma$ are significantly larger than the value of $19.35$~MeV/fm$^2$ from the quark-meson model~\citep{Hol18} but seem to agree with the larger value of $145 - 165$~MeV/fm$^2$ obtained within the NJL model~\citep{Lug13}.
In addition,
we note a recent work~\citep{Lug21} based on the MIT bag
model suggests that the quark vector interactions, which have been considered in the present CQM model, can strongly enhance the surface tension and the curvature energy of quark matter drops.
To self-consistently calculate the mass of $ud$lets and the surface tension of $ud$QM within the CQM model with the Hamiltonian Eq.~(\ref{Eq:Hami}) is challenging and deserves further investigation in future.
Moreover, it will be extremely interesting to explore the $ud$let production in various nuclear reactions related to the synthesis of superheavy nuclei, high-energy collider, ultrahigh-energy cosmic rays, and the nucleosynthesis in supernova explosions or binary QS merging. Compared to the three-flavor strangelets~\citep{Far84}, the two-flavor $ud$lets would be much easily produced in these nuclear reactions involving a large number of baryons ($\gtrsim 300$).

\section{Conclusion}
\label{Sec4}
Within the CQM model, we have demonstrated the recently reported GW190814's secondary component with mass $2.59^{+0.08}_{-0.09}M_\odot$ is likely to be a $ud$QS, supporting the hypothesis that the $ud$QM is more stable than nuclear matter and SQM.

\begin{acknowledgments}
This work was supported in part by the National Natural Science Foundation of China under Grant Nos. 12235010, 11975132 and 11625521, National SKA Program of China No. 2020SKA0120300,
the Shandong Provincial Natural Science Foundation, China (ZR2019YQ01), the Program for Professor of Special Appointment (Eastern
Scholar) at Shanghai Institutions of Higher Learning, and Key Laboratory for Particle Physics, Astrophysics and Cosmology, MOE, China.
\end{acknowledgments}

\end{document}